\documentclass[conference]{IEEEtran}
\IEEEoverridecommandlockouts
\newcommand{\etal}{{\it et~al. }}
\newcommand{\ie}{{\it i.e., }}
\newcommand{\eg}{{\it e.g., }}

\usepackage{optidef}
\usepackage[ruled,vlined]{algorithm2e}

\usepackage{cite}
\usepackage{amsmath,amssymb,amsfonts}
 \usepackage{algorithmic}
\usepackage{graphicx}
\usepackage{textcomp}
\usepackage{xcolor}
\def\BibTeX{{\rm B\kern-.05em{\sc i\kern-.025em b}\kern-.08em
    T\kern-.1667em\lower.7ex\hbox{E}\kern-.125emX}}
\begin{document}

\title{A Multi-Level Approach for Class Imbalance Problem in Federated Learning for Remote Industry~4.0 Applications}

\author{{Razin Farhan Hussain, Mohsen Amini Salehi}\\
{Email:\{razinfarhan.hussain1, mohsen.aminisalehi\}@louisiana.edu}\\
High Performance Cloud Computing (HPCC) Laboratory,\\ School of Computing and Informatics,
University of Louisiana at Lafayette, Lafayette, LA, USA\\
}

\maketitle

\begin{abstract}
Deep neural network (DNN) models are effective solutions for industry 4.0 applications (\eg oil spill detection, fire detection, anomaly detection). However, training a DNN network model needs a considerable amount of data collected from various sources and transferred to the central cloud server that can be expensive and sensitive to privacy. For instance, in the remote offshore oil field where network connectivity is vulnerable, a federated fog environment can be a potential computing platform. Hence it is feasible to perform computation within the federation. On the contrary, performing a DNN model training using fog systems poses a security issue that the federated learning (FL) technique can resolve. In this case, the new challenge is the class imbalance problem that can be inherited in local data sets and can degrade the performance of the global model. Therefore, FL training needs to be performed considering the class imbalance problem locally. In addition, an efficient technique to select the relevant worker model needs to be adopted at the global level to increase the robustness of the global model. Accordingly, we utilize one of the suitable loss functions addressing the class imbalance in workers at the local level. In addition, we employ a dynamic threshold mechanism with user-defined worker's weight to efficiently select workers for aggregation that improve the global model's robustness. Finally, we perform an extensive empirical evaluation to explore the benefits of our solution and find up to 3-5\% performance improvement than baseline federated learning methods.
\end{abstract}

\section{Introduction}
The rise of Industry 4.0 \cite{ccinar2020machine} elevates the utilization of smart IoT devices (\eg sensors, actuators, camera, routers) and advanced computing platform (\eg edge, fog computing) for deep neural network (DNN) applications (\eg oil spill detection, fire detection, toxic gas detection) in various industrial sectors (\eg smart oil field, smart farms, smart factory). For instance, in a remote offshore hydrocarbon reservoir, various oil and gas (O\&G) companies can have extraction sites nearby \cite{hussain2022iot,HUSSAIN2024479,hpcc20razin} that collect data using smart IoT devices. Here various companies' data acquisition fog devices can be a potential source of training data for deep neural network models. Hence, the network connectivity to mainland cloud data centers is unreliable and costly (\ie using satellite communication) \cite{greenrazin19,ums23}. For this reason, the O\&G companies typically rely on local computing systems (\ie micro or mini data centers a.k.a fog) for their computing demands \cite{hussain2019federated}. To collect data (\eg image, video, audio) for various machine learning (ML) and DNN applications \cite{cadavid2020machine}, companies utilize different mediums (\eg satellite, drone, aircraft) that are expensive and area-dependent. For instance, each region has its physical characteristics, and therefore they are exposed to different data sources. Here the main objective of the company is to improve the DNN model without sharing their private data.

\begin{figure}
\centering
\includegraphics[width=0.5\textwidth]{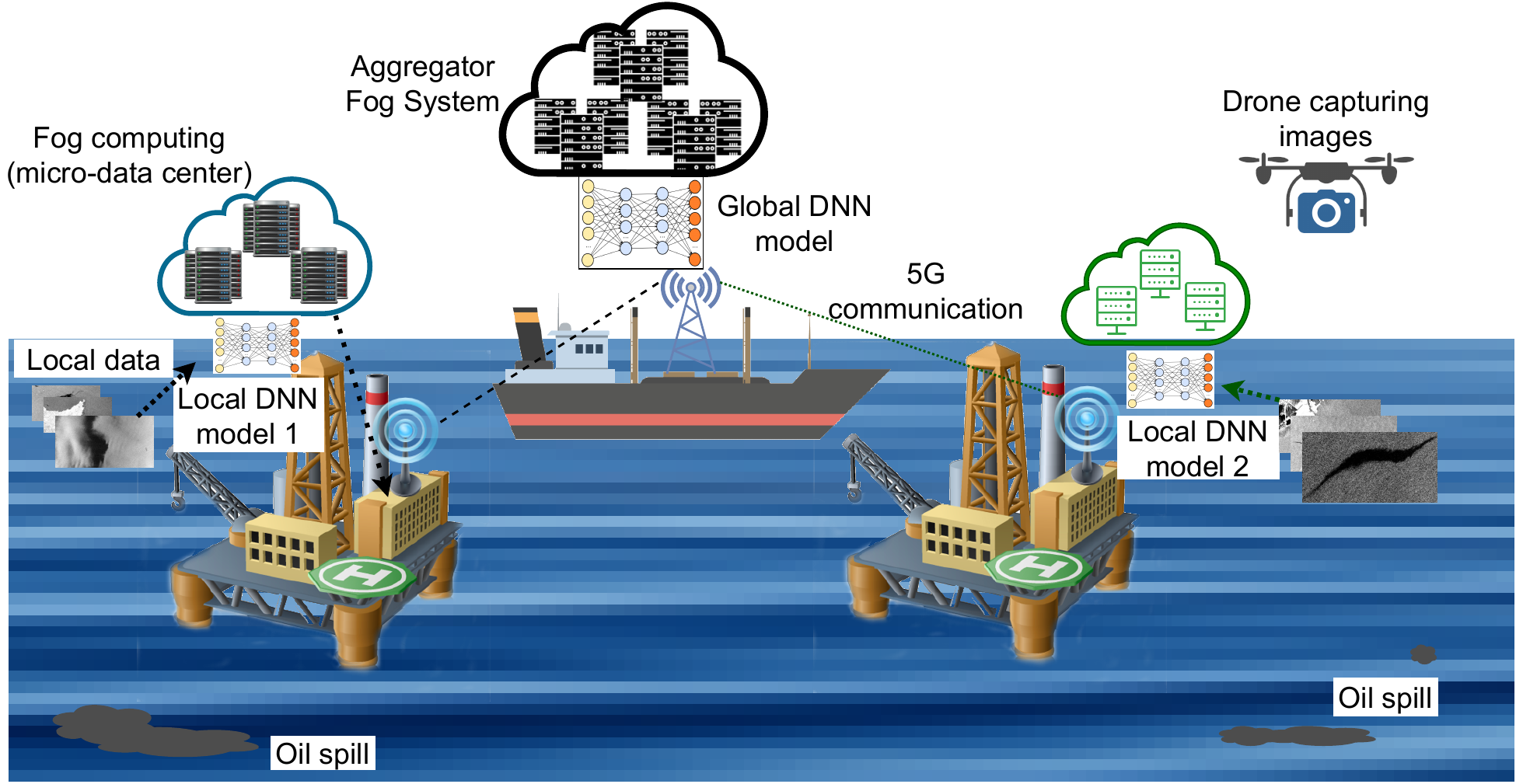}
\caption{A federated learning setup in fog federation. Multiple company share their fog systems to train oil spill detection DNN model where data security is preserved by federated learning.}
\label{fig:fogFLintro}
\centering
\end{figure}

Considering an oil spill incident in remote offshore oil field, our motivation is to train an oil spill detection DNN model utilizing the fog computing resources available in offshore. The main challenge with conventional centralized machine learning \cite{hossen2020object} is that the security concern often leads to a less accurate DNN model due to a lack of training data \cite{hpcc20razin}. Hence, rival companies may not want to share their private data \cite{singhai2021investigation,zob22}. Moreover, the oil spill is a low probability event that can appear in the training image rarely. Consequently, the training data set may have fewer occurrences of oil spill classes, making the problem context-dependent. Therefore, O\&G companies are motivated to adopt a solution to improve their DNN model's quality without increasing data collection costs and preserving data privacy. Hence, we propose to utilize O\&G companies' local computing resources (\ie fog computing) by forming a federation of computing systems that solve unreliable and costly network communication to mainland cloud data centers and incorporate more data for training.

\begin{figure}[!t]
\centering
\includegraphics[width=0.5\textwidth]{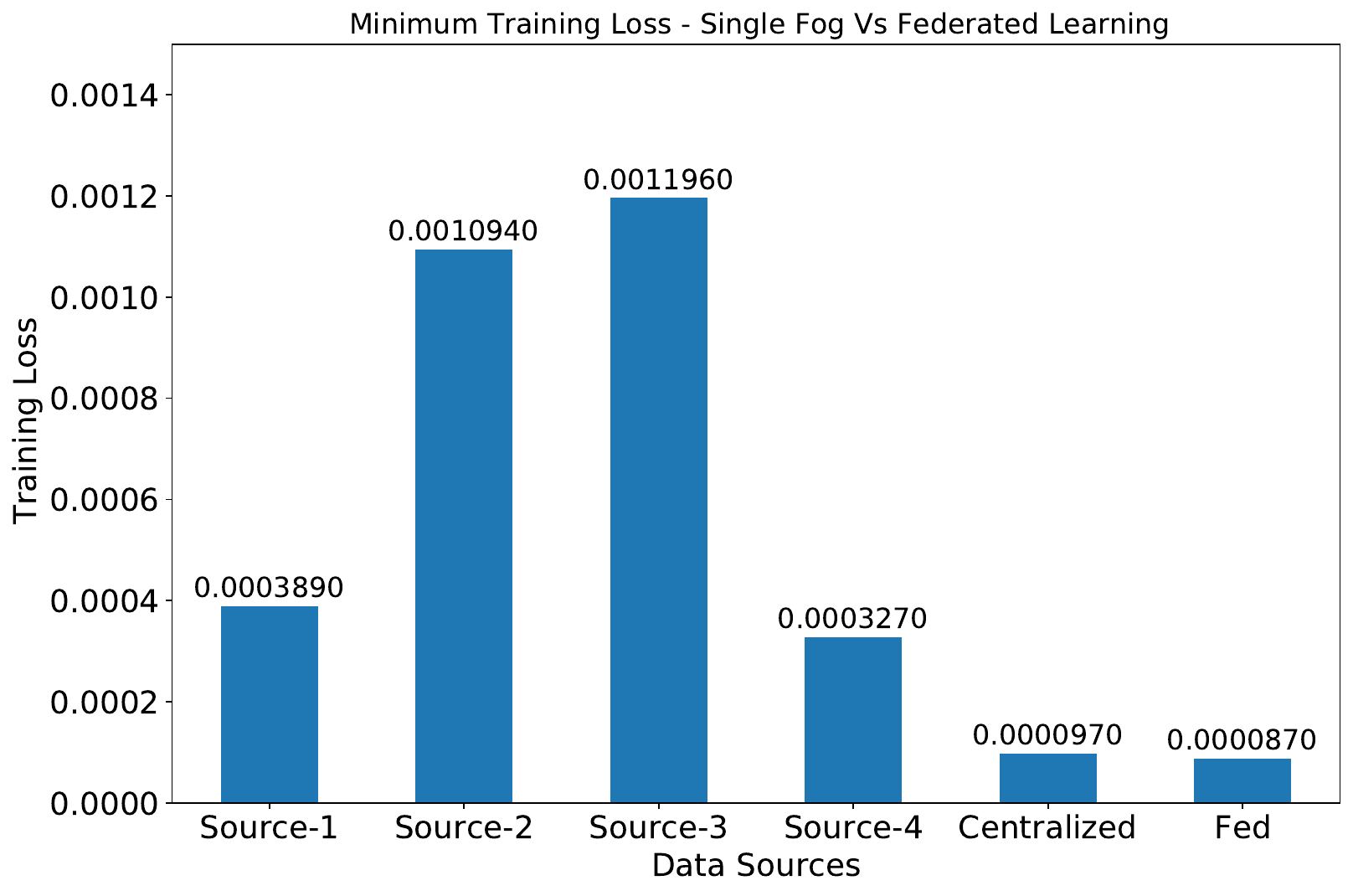}
\caption{Comparison of minimum training loss for single machine (fog) learning and federated learning. The first four bars from the left represents single machine learning with four different data sources (S1, S2, S3, and S4), fifth bar is single machine learning with combined data source as centralized learning, and right most bar represent the federated learning. Minimum training loss for all cases are considered, and federated learning has the minimum loss comparing other single machine learning.}
\label{fig:FLnonFl}
\centering
\end{figure}

\begin{figure}[!t]
\centering
\includegraphics[width=0.5\textwidth]{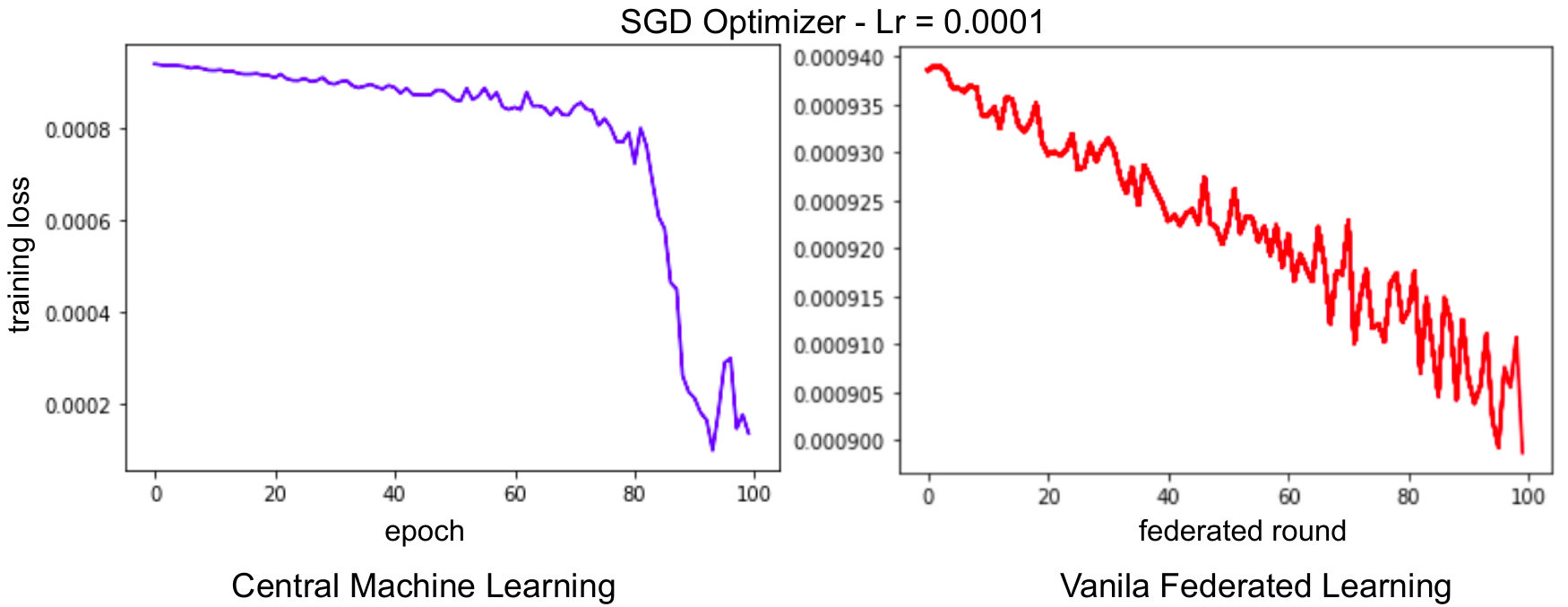}
\caption{The training loss of Unet model with oil spill data set in both central machine learning, and federated learning setup. The convergence trends of training losses are similar in both setup.}
\label{fig:flconvergence}
\centering
\end{figure}

One of the viable solutions (presented in figure \ref{fig:fogFLintro}) for security issues can be federated learning (FL) \cite{zhou2020privacy,fedZehui} that also support the utilization of computing resources across the fog federation. However, the problem is that data come from various sources, and it is feasible to assume that data distribution tends to be non-identical and independent distribution (non-IID). As such, lack of any priority class (\ie \textit{oil spill}) that is termed as \emph{class imbalance} \cite{duan2020self} can reduce the performance of the global model in an FL setup. Hence, ignoring the class imbalance issue, current federated learning methods \cite{mcmahan2017communication} are providing less robust DNN models. The robustness of FL training mainly depends on the local workers' aggregation that affects the global model's performance. In this case, a DNN detection model can show misleading high accuracy for all other classes while providing low performance for the desired class (oil spill). As such, we train a DNN model with an imbalanced data set using federated learning set up to understand the benefits of FL on class imbalance issues. We compared the training loss with the conventional method of single machine learning. In the figure \ref{fig:FLnonFl}, the x-axis represents the various data sources (\ie S1, S2, S3, S4, and centralized) with lowest loss epoch, whereas the y-axis represents the training loss. The centralized data source combines all four data sources with single central machine learning. The result (depicted in figure \ref{fig:FLnonFl}) indicates that federated learning has the lowest loss comparing single machine learning. In addition, to verify the convergence of the DNN model in federated learning setup, we capture the training loss for both central machine learning and vanilla federated learning setup. Figure \ref{fig:flconvergence}, represents the similar convergence trend for federated learning setup. Therefore, we are motivated to utilize a federated learning setup for oil spill detection DNN model training.

Among two types of federated learning (synchronize and asynchronous), we propose to utilize the synchronize FL method as it is a proven model, especially for class imbalance issues \cite{duan2019astraea,mhaisen2020analysis}. To overcome the challenges of FL for oil spill detection, we have adopted an objective function (loss function) to train the local model considering the class imbalance problem. Furthermore, considering the priority class (\ie oil spill), we introduce a weight for each participating worker that reflects the performance of the global model over class imbalanced suffer class (\ie oil spill). Finally, a dynamic threshold mechanism has been proposed to select relevant workers efficiently considering the global model's performance. Therefore, the contribution of this work can be listed as follows:  
\begin{itemize}
    \item Addressing the impacts of class imbalance issue that is inherited in training data.
    \item Proposing a weight of FL worker that implicitly reduces the effect of class imbalance in the global model.
    \item Introducing a dynamic threshold mechanism that ensures the global models' robustness with relevant federated workers.
\end{itemize}

The rest of the paper is organized in the following manner. Section \ref{relatedwork} presents the related work and background of federated learning. Section \ref{problemForm} demonstrate the problem formulation, and section \ref{sysMod} describes the system model. Section \ref{solution} represents the solution method. In section \ref{expSetup}, the environmental setup is explained whereas section \ref{performanceEval} explains the performance evaluation, and experimental results. Finally, section~\ref{concl} concludes the paper with a discussion and future avenues for exploration.

\section{Related Work \& Background of Federated Learning}\label{relatedwork}
Federated learning has drawn significant attention to the research community due to its security improvement and distributed nature. Especially, with the edge \cite{ye2020federated}, and fog computing \cite{veillon2019f} platforms utilizing the federation, has been the best suitable candidate for many remote industries. However, most of the current research mainly focuses on improving federated learning framework either concerning communication or considering computing resource bottleneck problems with the same existing data sets (\eg MNIST, CIFAR) and the same area of image classification. Hence, current research has yet to explore other relatively new data sets and DNN areas (object detection, semantic segmentation) that may cause problems in model improvement techniques.

The Federated learning setup provides security as the training data does not leave the owner's device. However, confidential information can still be exposed to some extent from adversaries' analysis on the differences of related parameters trained and uploaded by the clients, \eg weights trained in neural networks \cite{shokri2015privacy,ma2020safeguarding, wang2019beyond}. Truex \etal \cite{truex2019hybrid} combine both secure multiparty computation and differential privacy for privacy-preserving FL. They use differential privacy to inject noises into the local updates. Then the noisy updates will be encrypted using the ``paillier cryptosystem'' \cite{paillier1999public} before being sent to the central server.

The federated learning, as an active research area, brings many potential improvements in computer vision \cite{liu2020fedvision,pillutla2019robust, liu2020experiments}. Liu \etal in \cite{liu2020experiments} performed federated learning for identifying covid-19 patients from chest X-ray images utilizing various hospitals patients data record. Due to preserving security, the FL model provides an effective solution to address covid-19 patients from X-ray images. Another application of federated learning could be oil spill detection using satellite image data set that usually fall under binary classification problems. However, very few research works have been conducted on semantic segmentation for oil spill detection. Moreover, multiple classes can be present in the same image, and each class is labeled with a specific color. Especially, using federated learning in semantic segmentation problems is rare in the academic research community.

Deep learning network models (\eg unet, linknet, fcn-8) for semantic segmentation need extensive data to train a robust model. However, acquiring adequate data for training in the medical image domain is difficult due to security, privacy, and data-ownership challenges. To overcome these challenges, Sheller \etal in their work \cite{sheller2018multi} first introduced federated learning in semantic segmentation for brain tumor identification. In this work, authors utilize the semantic segmentation model (Unet) on multimodal brain scans and empirically discovered similar performance for those models trained by sharing data. However, the class imbalance is not present in their work. They also did not consider the robustness of the global model with an effective client selection mechanism. Our work would be one of the first pieces of research to utilize federated learning in oil spill detection that considers the class imbalance issue in the global model to the best of our knowledge.

Additionally, we address the worker selection problem that affects the robustness of the global model. Hence, from the system administration point of view, adequate consideration of relevant workers can enhance the global model's accuracy in terms of semantic segmentation metric named mean intersection over union (mIoU). Therefore, our primary focus in this work is to increase the global model`s accuracy (\ie mIoU) with a higher probability of identifying oil spill class.

\section{Problem Formulation}\label{problemForm}
The oil spill detection problem can be well defined in the semantic segmentation domain of deep learning. Hence we consider a satellite image data set \cite{krestenitis2019oil} consisting of five classes, namely oil-spill, look-alike, land, ship, and sea-surface. Here, each class is labeled as an individual color in the ground truth image. For training, a deep neural network (DNN) model (\eg Unet) with federated learning settings a set of workers $S = {1,2,3,...,S}$ are considered with its local data set $D^L$ where $L \in S$ with ~$n_L$ samples. Here, $D= \bigcup_{L \in S} D^L$ is the full training data set. The total size of these workers' data set for a random set of workers $S^\prime$ is $N(S^\prime) = \sum_{\mathcal{L} \in S^\prime}n_L $. The objective loss function over a model \textit{m} and a sample \textit{z} can be denoted as $\mathcal{L}(m,z)$.

\begin{figure}[h!]
\centering
\includegraphics[width=0.50\textwidth]{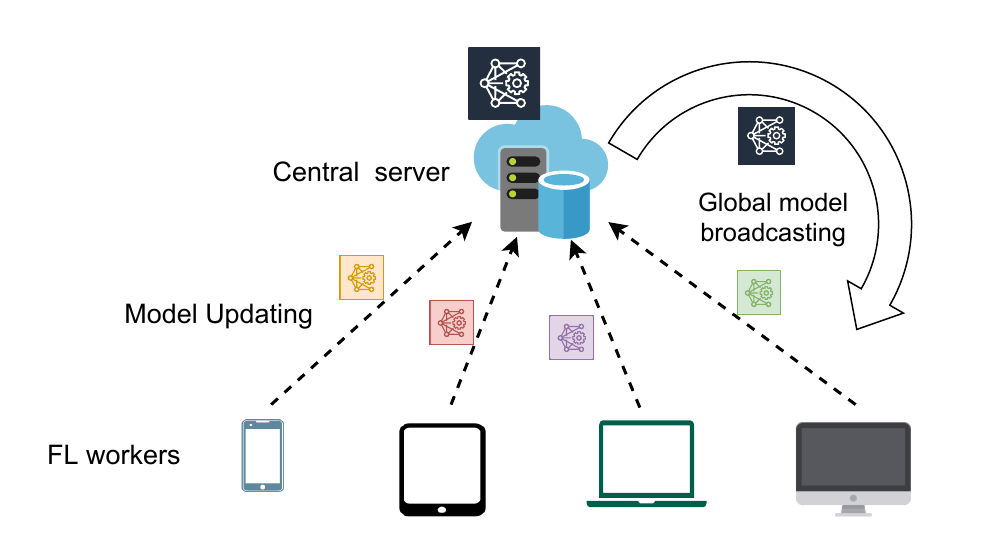}
\caption{A typical federated learning (FL) setup with different worker clients participating in training phase. Global model is broadcast or downloaded to the worker clients. Then workers train the global model with their local data and send the updated model to the central server for updating the global model.}
\label{fig:typicalFL}
\centering
\end{figure}Then in most prior FL work, the goal is to solve the following
\begin{mini*}|s|
{w}{f(w) = \sum_{m=1}^{M} p_m F_{m}(w)}
{}{}
\end{mini*}
where $p_m = \frac{n_m}{D}$ is the fraction of the total data worker, and thus, $\sum_{m}p_m = 1$. The local objective $F_m$ of model $m$ is typically
defined by the empirical loss over local data, $F_m (w) = \frac{1}{n_m} \sum_{j=1}^{n_m}\mathcal{L}_{j}(m,z)$. Here, \textit{w} is the model parameters that are used for predicting loss over a sample data, and the goal is to find the optimal $w$ for which the loss should be minimized. Accordingly, we focus on utilizing a loss function that consider class imbalance problem in local data samples, and select a set of client worker's models to aggregate that have certain level of accuracy (\ie mean intersection over union (mIoU) for semantic segmentation). Hence, our new objective for this work would be as following:
\begin{mini*}|s|
  {w}{f(w) = \sum_{m=1}^{M} p_m F_{m}(w)}
  {}{}
  \addConstraint{mIoU(m)}{>=\gamma}
  \addConstraint{\theta}{>1}
\end{mini*}
Where, $\gamma$ is a dynamic threshold (initial value set to 50\% or 0.5) for checking the local trained model's mIoU estimated with auxiliary test data, and $\theta$ is the user defined worker's weight with respect to oil spill class. Both of this parameters are used to select the relevant worker's model for aggregation into the global model that ensure the robustness, and consistent convergence of the global model.
\begin{table}[]
\centering
\caption{Pixel distribution for each of the class in oil spill detection data set}
\begin{tabular}{|l|l|}
\hline
Class       & Pixels  \\ \hline
Sea Surface & 797.7 M \\ \hline
Oil Spill   & 9.1 M   \\ \hline
Look-alike  & 50.4 M  \\ \hline
Ship        & 0.3 M   \\ \hline
Land        & 45.7 M  \\ \hline
\end{tabular}

\label{tab:classImb}
\end{table}

\section{System Model}\label{sysMod}
An oil spill is a rare event in the satellite image data set, being a low probability event. Hence, this low occurrence event creates the class imbalance issue in oil spill detection training, leading to a less robust model against oil spill class or other victim classes. In the oil spill data set that we collect from MKLab \cite{krestenitis2019oil,krestenitis2019early}, a computer vision research institute in Greece, the class imbalance issue is found significantly for oil spill class and ship class (table \ref{tab:classImb}). In addition, typically, oil spill detection DNN model's evaluation metrics can be misleading. For instance, DNN model accuracy considering classification problems for oil spill detection (when the data set has only two classes \eg oil spill or not) can be significantly high. However, the intersection over union (a.k.a IoU) can be unsatisfactory for any particular class (\eg oil spill). The low performance is that the DNN model is mostly exposed to the background class, and the oil spill class suffers from class imbalance. Hence, the problem of oil spill detection needs to be considered according to the data set, and specialized evaluation metrics need to be in place to train a model. Therefore, we consider oil spill detection as a semantic segmentation \cite{yekeen2020novel} problem and utilize intersection over union as the evaluation metric.  
We assume that class imbalance results from local data samples that influence the local FL worker models and gradually the global model. Hence the idea is to utilize a loss function that considers class imbalance issues. To achieve this goal, we choose the tversky index \cite{abraham2019novel} to use as our objective loss function for training the DNN model that can be defined as follows:
\begin{equation}
T_i = \frac{TP}{TP+\alpha*FN+\beta*FP}
\label{tverskyIndex}
\end{equation}
Where, $TP$ = True Positive, $FN$ = False Negative, $FP$ = False Positive, $\alpha$ = penalizing factor for false negative prediction, and $\beta$ = penalizing factor for false positive prediction. The training with worker's local data is considered as local level where \emph{tversky loss} ($tversky loss = 1 - T_i$) \cite{salehi2017tversky} function is used to penalize the false negative prediction. We choose the tversky loss function due to its customization option (\eg $\alpha$,$\beta$) for penalizing false-negative prediction of the model. Using empirical experiments, we select the best parameter value of the tversky loss function for our solution.


At the global level, a significant impact can be made with the worker selection technique. For instance, if the worker model is trained with a local data set that has less occurrence of oil spill class, then adding this update to the global model would reduce the model accuracy for the oil spill class. Therefore, typically all the local trained worker models are averaged (FedAvg\cite{mcmahan2017communication}) and set to the global model. Although other aggregation algorithms (\eg FedSGD, FedProx) are present in literature, the FedAvg has proven performance against class imbalance issues. Accordingly, we start working with the FedAvg algorithm and utilize our system solution on top of FedAvg to make the algorithm robust against the class imbalance issue. 
We utilize class-level intersection over union (IoU) to estimate worker weight and mean IoU as threshold estimate for selecting relevant workers. The worker's weight can be defined as follows: 
\begin{equation}
\theta_i = \frac{IoU(oil~spill~ class)}{mIou(M_i)}
\label{workerWeight}
\end{equation}
Where $\theta_i$ is the weight of worker $i$, and $M_i$ is the trained model of worker $i$. We consider worker weight ($\theta$) greater than 1 for selecting relevant local trained worker models. The reason is that those workers have a significant performance for our priority class (\ie oil spill class). The priority class can be different for a different context, and accordingly, weight can be adjusted. The relevant worker selection could lead to having fewer workers model for aggregation and decrease the performance of the global model. To handle this issue, we dynamically adjust the threshold value after every federated round depending on the number of relevant workers selected that maintains the robustness of the global model throughout the federated rounds.

\begin{figure}[!htbp]
\centering
\includegraphics[width=0.47\textwidth]{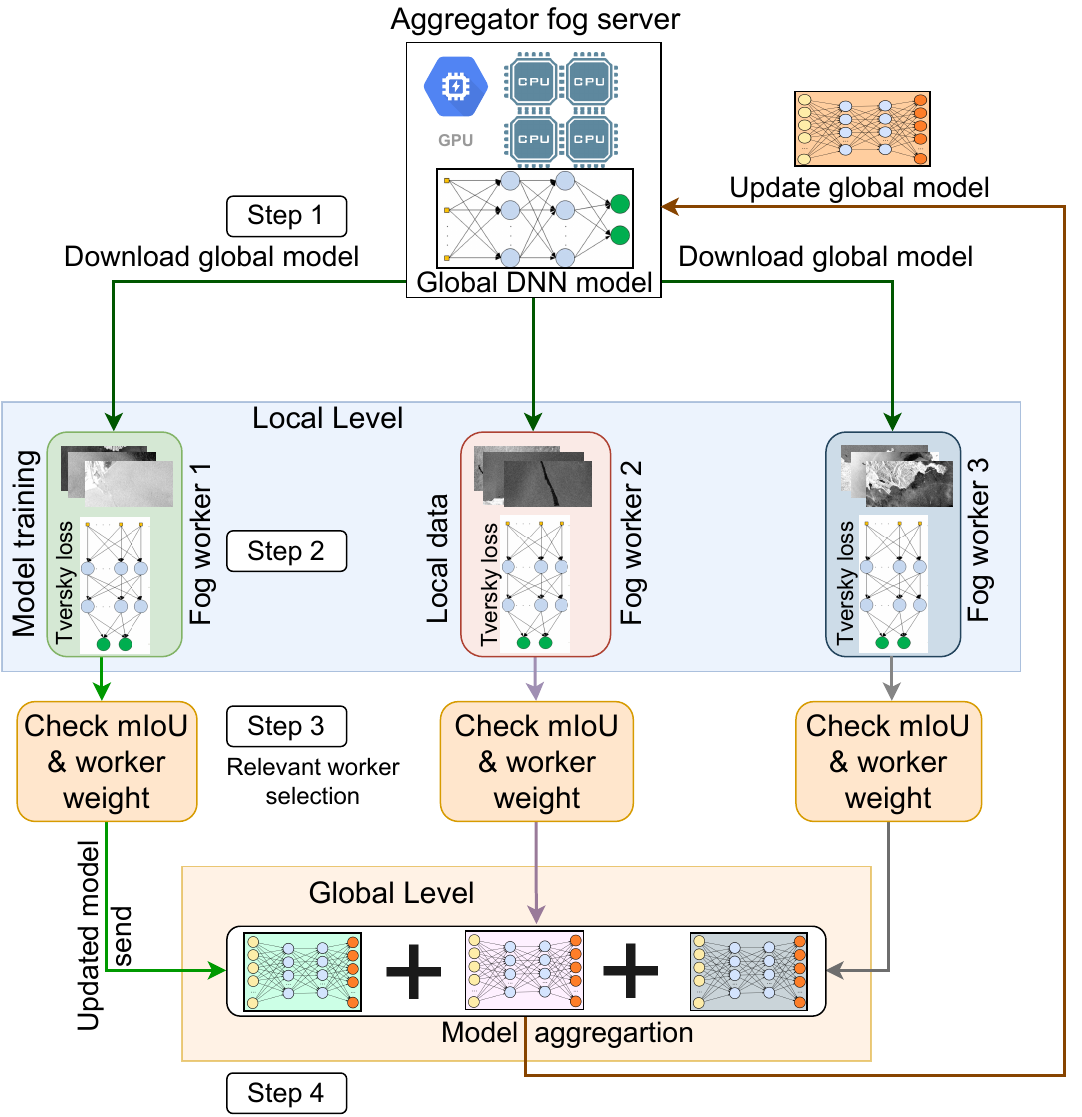}
\caption{Federated learning training considering class imbalance and global convergence. Tversky loss is used in the training considering class imbalance. After training of each epoch, mean intersection over union (mIoU) is checked with a dynamic threshold for global convergence.}
\label{fig:solFL}
\centering
\end{figure}

\section{Federated Learning Solution to Reduce the Class Imbalance}\label{solution}
In a typical federated learning setup, the server (\eg fog device, cloud) stores a global ML model that is broadcast to the participating workers for training. We use one of the popular semantic segmentation DNN models named Unet \cite{ronneberger2015u} for oil spill detection in the FL setup. The figure \ref{fig:solFL} represents a pictorial view of our solution. At first, some fog nodes agree to participate in the FL training, and they download the global model from the fog server presented in step 1 of figure \ref{fig:solFL}. Then, downloaded ML models are trained with their local data using the Tversky loss function in step 2. In step 3, local models are checked to select relevant workers (\ie presented on algorithm \ref{relWorkSel}) for aggregation in the global model. Finally, in step 4, selected workers updated models are aggregated (new model), and the previous global model is updated accordingly. This whole process is considered a \emph{federated round}. The updated model is again downloaded by participating workers for the next federated round, and the training continues. Finally, the proposed solution is presented in algorithm \ref{fedbal}.
 

\begin{algorithm}[!h]
   Initialize global model, $m_g$, relevant Worker List, $r_f$, not relevant worker list, $n_f$, Initial threshold, $t_h = 0.50$

   \For{each federated round f = 1,2,...}{ 
        m $\leftarrow$ max(C.K,1)
        
        $S_t \leftarrow$ random~set~of~m~workers

        \For{each worker $k \in S_t$ \textbf{in parallel}}{
              ClientUpdate(k,$m_g$) 
              
       /*\texttt{training with local data}*/
    }
      $r_f, n_f$ = relevantWorkerSelection($S_t$,test\_data,$t_h$)
   
     $t_h$ = dynamicThreshold($r_f,n_f,S_t$)
   
     $m_g$ = aggregateModels($m_g, r_f$)    
   
   }
\caption{FedBal. K is total workers pool where $k$ is individual worker, and C is the initial worker selection percentage}\label{fedbal}
\end{algorithm}

Our proposed (\emph{FedBal}) algorithm starts with initializing global model $m_g$, relevant worker list, $r_f$, not relevant worker list $n_f$, and setting the threshold, $t_h$ value to 0.50 (Average global models accuracy in terms of mIoU is 50\%). The federated round continues as a for loop that is presented with variable $f$. Then $m$ number of workers are selected from $K$ participating worker, and assigned to selected worker list, $S_t$ for training (``ClientUpdate'' function) with their local data. After the local training, relevant workers ($r_f$) are selected using function ``relevantWorkerSelection'' (presented in algorithm \ref{relWorkSel}). According to the relevant workers' list ($r_f$), the threshold is adjusted for the next federated round with the function, ``dynamicThreshold'' that is presented in algorithm \ref{dynamicThreshold}. Finally ``aggregateModels'' function aggregates the local trained models of the workers into the new global model, $m_g$.

\begin{algorithm}[!h]
\DontPrintSemicolon
  \textbf{Input:} Selected workers for training ($S_t$), test data ($test\_data$), threshold ($t_h$)\\
 \textbf{Output: } Relevant workers for aggregation ($r_f$)
  
  \For{each worker, $w$ in $S_t$}    
    { 
        $mIoU(w)$ = Estimate mIoU with $test\_data$\\
        $\theta_w$ = Measure worker weight
        
        \If{$t_h$ $\leq$ $w(mIoU)$ and $\theta_w$ $\geq$ 1}
        {
            $r_f$ $\leftarrow$ $w$
        }
        \Else
        {
    	   $n_f$ $\leftarrow$ $w$ 
        }
    } 

   return $r_f$, $n_f$
\caption{Relevant Worker Selection}\label{relWorkSel}
\end{algorithm}

The proposed solution's global level is triggered in the ``relevantWorkerSelection'' function, where trained worker models are evaluated according to their weight, $\theta$, and $mIoU$ value. When the local trained model has a higher mIoU than the threshold ($t_h$) value, and weight ($\theta_w$) is greater than ``1'', the worker is considered as relevant and added to the relevant worker list ($r_f$). Otherwise, the worker is added to the not relevant worker list ($n_f$). Both of the lists are utilized to determine the threshold ($t_h$) value by the ``dynamicThreshold'' function. 

\begin{algorithm}[!h]
\DontPrintSemicolon
  \textbf{Input:} $r_f$,$n_f$, $S_t$\\
  \textbf{Output: }  New threshold, $t_h$

    \If{length($r_f$) is 25\% of $S_t$}
    {
    	$t_h$ = Highest mIoU of $n_f$
    }
    \ElseIf{length($r_f$) == 0 or length($r_f$) is 50\% of $S_t$}
    {
    	$t_h$ = Median mIoU of $n_f$ 
    } 
    \Else
    {
        $t_h$ = $t_h$ + 0.01
        \tcc{Increase threshold to improve the accuracy of global model}
    }
   return $t_h$

\caption{Dynamic Threshold Selection}\label{dynamicThreshold}
\end{algorithm}

The dynamic threshold mechanism presented in algorithm \ref{dynamicThreshold} ensures the robustness of the global model by adjusting the $t_h$ value in each federated round. This algorithm checks the length of $r_f$ to increase or decrease the initial threshold value. When relevant workers are 25\% of participating workers ($S_t$), then the highest mIoU of the not relevant workers is set as a new threshold value. Similarly, when length of $r_f$ is zero or 50\% of $S_t$ list, the median mIoU value of $n_f$ list is considered as new threshold value. In every other case, the threshold value is increased by 0.01 to improve the global model's performance. In this way, in every federated round,$f$, the global model updates and converges to a model that is robust against class imbalance with guaranteeing performance for our priority class, oil spill.


\section{Environmental Setup for FedBalance Method }\label{expSetup}
The Federated learning setup can be synthesized by PyTorch's one of the popular library pysyft \cite{ziller2021pysyft}. Due to pysyft's customization capability, we develop our solution using pysyft. This work considers oil spill detection as a semantic segmentation problem and utilizes a real-world satellite SAR image data set to train the DNN model. To execute the DNN training operation, we used Google's Colab \cite{bisong2019google} run-time environment that provides a GPU platform with a high-speed ram of size 24 GB with storage of 128 GB.

The Colab provides Tesla P100, T4, or similar GPUs for the paid ``pro'' version. We utilize pysyft's virtual worker's concept to synthesize fog devices. Our primary focus in this work is to reduce the impact of class imbalance on the global model and ensure robust global model training. Hence we concentrate on the computation part of FL and ignore the communication (\ie network) of conventional FL setup. Our federated learning setup can be utilized for any aggregation algorithms (\eg FedAvg, FedSGD, FedProx), and as such, we develop our codebase on top of these baseline algorithms.

To explore the benefits of our solution, we executed 50 rounds of federated learning for three of the FL techniques (\eg FedAvg, FedProx, FedSGD) and compared the global model`s mean intersection over union (mIoU) with our proposed solution. In most of our experimental results, we use the first 20 federated rounds to reflect the differences between our proposed method and baselines. The reason behind choosing these values for the training parameters (\eg number of epoch, number of federated rounds) is to observe a significant difference among the aggregation algorithms.

\vspace{-4mm}
\begin{figure}[h!]
\centering
\includegraphics[width=0.45\textwidth]{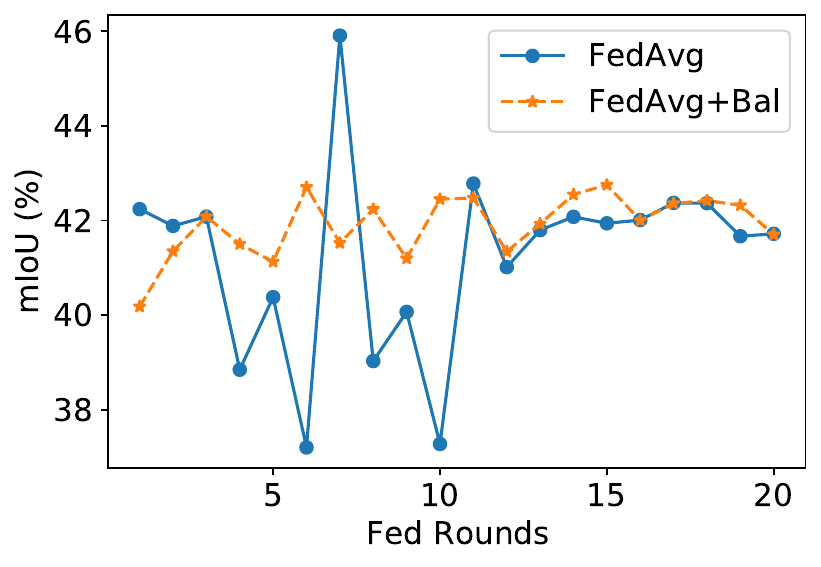}
\caption{The performance comparison of global models in terms of mIoU using FedAvg and FedBal methods. The data distribution is non-IID, the number of workers are 6, and in each fed round 50 epochs of training has been performed.}
\label{fig:noniid}
\centering
\end{figure}


\begin{figure*}[h!]
\centering
\includegraphics[width=\textwidth]{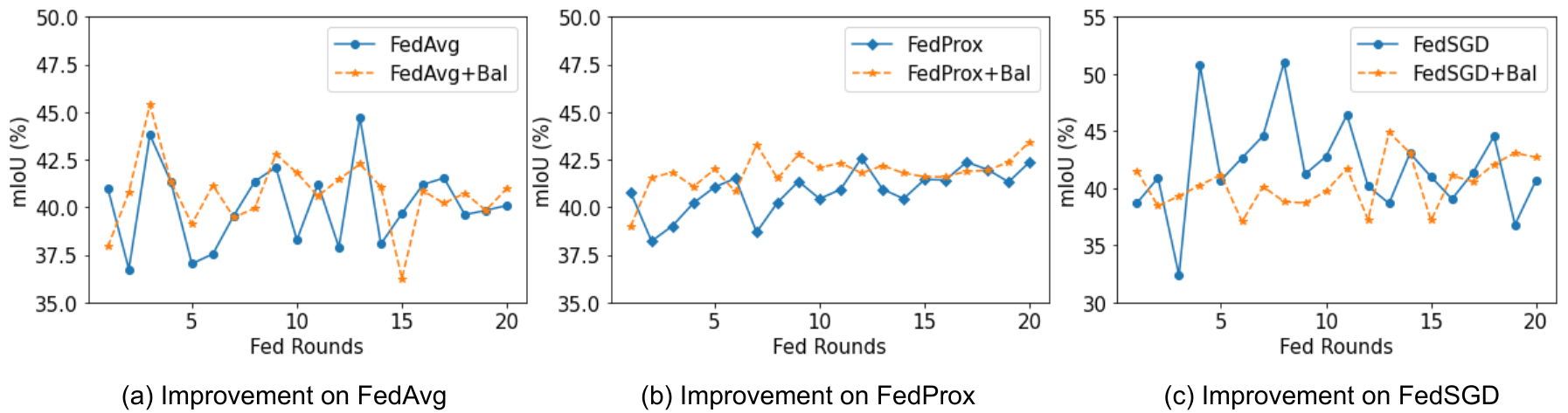}
\caption{Comparison of proposed method's performance improvement on FedAvg, FedProx, and FedSGD respectively in non-IID and unbalanced data distribution.}
\label{fig:noniidunbalance}
\centering
\end{figure*}


\section{Performance Evaluation of FedBal Method}\label{performanceEval}
The federated learning setup is always beneficial for fog devices where data tends to be generated frequently. To understand the advantage of using our proposed FL setup, we first tune the tversky loss function to find the minimum loss and then utilize the optimum configuration for the rest of the experiments. As we focus on class imbalance issues, we utilize non-IID and non-IID \& unbalance data distribution to measure the global model's accuracy metric (mIoU) throughout the federated rounds. The main difference between the above mention setup is the number of classes distributed to the federated workers. In non-IID distribution, all the participating FL workers get the same number of classes, although classes are different for each worker. On the contrary, in non-IID \& unbalance distribution, each participating worker gets a different number of classes. We consider following the data distribution setup of McMahan's \cite{mcmahan2017communication} work that validates our experiments.




\subsection{Tuning Loss Function for Minimum Loss}
To find the minimum loss from the tversky function, we change the alpha parameter value from 0.6 to 0.8 and capture each training epoch's loss. The main goal is to find the optimum alpha value for which the loss will be minimal. From this experiment, we find that for the alpha value of 0.7, FedBal has minimum training loss. However, when we increase the alpha value to 0.8, the training loss does not decrease, which means we can penalize false negatives up to a certain point (\ie $\alpha = 0.7$). The reason is that while we are penalizing false negatives, the false positive predictions are ignored (\ie $\alpha + \beta = 1 $) as well. Hence for a higher value of alpha, we get less benefit by penalizing false-negative predictions. Therefore, we use $\alpha = 0.7$ for the rest of our experiments throughout this work.

\subsection{Global Model Performance with non-IID Data Distribution}
In a real-world scenario, data distribution among FL workers is typically non-IID, where every worker will get a fixed number of classes for local training. Hence, we consider providing two classes for each worker, and these classes are different for every worker. We measure the mIoU of the global model for FedAvg and FedAvg+Bal, respectively, after each federated round. The result is provided in figure \ref{fig:noniid} for 20 federated rounds with six federated workers.

Figure \ref{fig:noniid} reflects that FedAvg+Bal has a consistent performance (mIoU) for 20 federated rounds. The FedAvg+Bal method has less uncertainty (fewer spikes in orange line of figure \ref{fig:noniid}) across the federated rounds. Although FedAvg+Bal has less significant performance improvement than FedAvg, the average mIoU of the global model of FedAvg+Bal is higher than FedAvg. This consistent performance of FedAvg+Bal represents the robustness of our method across the federated rounds.  

\subsection{Global Model Performance with non-IID and Unbalanced Data Distribution}
The non-IID and unbalanced data distribution mean each FL worker has a different number of classes. For instance, worker one can have two classes, whereas worker two can have only one class. Hence, we measure the mIoU of the global model and compare our method to the other three baseline methods named FedAvg, FedProx, and FedSGD, respectively. As our method is considered an improvement on any federated learning aggregation method (\eg FedAvg, FedProx, FedSGD), we compare the baselines separately in three different sub-figures. The result is demonstrated in figure \ref{fig:noniidunbalance} where the x-axis represents the federated rounds, and the y-axis represents the mean intersection over union (mIoU) of the global model.

Figure \ref{fig:noniidunbalance} (a) represents that our proposed method outperforms FedAvg, FedProx, and FedSGD respectively in the final round. Although, from the 17th to 19th federated round, FedSGD has better mIoU than FedSGD+Bal. The performance improvement for FedProx+Bal is significant for FedProx, due to the utilization of the left out workers in the global model. It is also visible that baseline methods have severe uncertainty (more spikes), whereas our method has comparatively consistent performance throughout the federated rounds. The main reason behind this consistency is the relevant worker selection with a dynamic threshold mechanism.

\subsection{FedBal's Performance on Class Imbalance Intensity}
\begin{figure}[h!]
\centering
\includegraphics[width=0.40\textwidth]{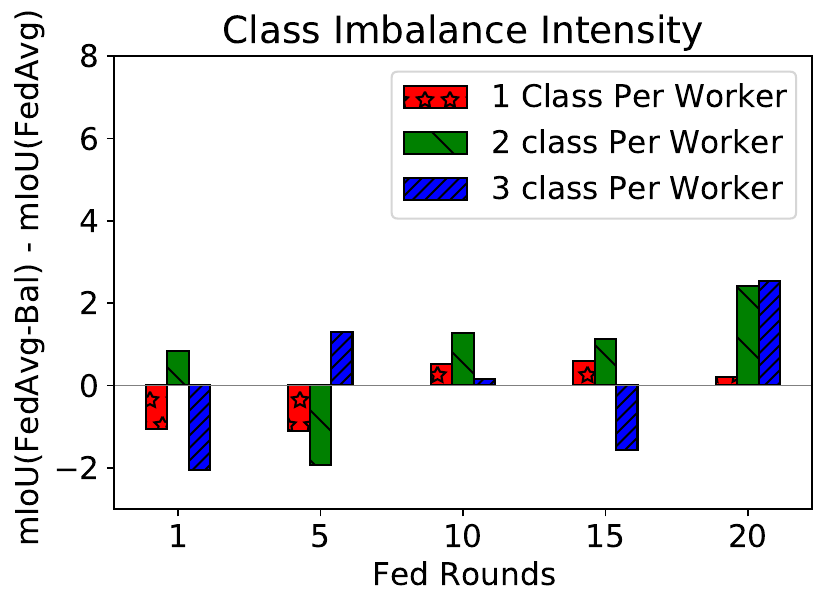}
\caption{The difference of mIoU of FedBal, and FedAvg is plotted as barchart for 3 cases of class imbalance intensity (1 class, 2 class, and 3 class).}
\label{fig:classimbalance}
\centering
\end{figure}

To explore the class imbalance intensity, we distribute the classes from high imbalance to low imbalance using a non-IID setup and measure the mIoU of the global model for FedAvg+Bal and FedAvg across the federated rounds. We estimate the difference of mIoU values for each federated round for three cases (one class, two classes, and three classes distribution) of imbalanced data distribution and plot a bar chart presented in figure \ref{fig:classimbalance}. Here positive values indicate FedAvg+Bal's improvement over FedAvg, and negative values represent the opposite.

Figure \ref{fig:classimbalance} represents the performance improvement of our method over FedAvg algorithm across 20 federated rounds of training. We plot the difference in every five federated rounds. For the first and fifth federated rounds, the improvement of FedAvg+Bal over FedAvg is not significant. In the 10th federated round, the difference values are all positive, and in the final federated round (20th), we find the highest performance of FedAvg+Bal over FedAvg. We also notice that for two classes per worker distribution, our method constantly outperforms FedAvg. For high intense class distribution (only 1 class per worker), FedAvg+Bal starts to perform well after the 10th round. In the final round, we find that for 3 class distribution FedAvg+Bal has a significant improvement. The main reason behind the less significant performance could be the dynamic threshold mechanism that starts with an average mIoU value (50\%) and dynamically change over federated rounds to increase the performance of the global model. After the 10th round, the threshold becomes stable with sufficient relevant workers, and we see performance improvement for the last ten rounds of federated training.   

\section{Conclusion}\label{concl}
The federated learning technique has revolutionized distributed machine learning, especially considering data privacy and computational flexibility. Federated learning brings the ML model to the data generation sources that is less expensive and secure than cloud data centers. Although federated learning can overcome security, and cost challenges, the class imbalance issue can degrade the global model's performance. As such, we focus on reducing the effect of class imbalance at the local level (\ie tversky loss) while training the model, and the global level (\ie relevant worker selection) while aggregating the federated worker. 

In the empirical evaluation we find that for non-IID setup FedBal demonstrates a consistent performance across the federated rounds. Although, FedBal's average performance for the non-IID setup is better than FedAvg. For non-IID and unbalanced setup, FedBal outperforms FedAvg, FedProx, and FedSGD respectively in the 20th federated round. In the class imbalance intensity experiment, we find FedBal performs better than FedAvg in the final federated round (20th round) in three of the cases. Although, for high-class imbalance (only one class per worker) intensity FedBal has less significant improvement (0.25\%) whereas for low-class imbalance (three class per worker) intensity FedBal shows significant performance improvement (more than 2\%). 
The ML training parameters (number of epochs, federated rounds, optimizer, batch size) can be tuned in a more granular way to explore the areas of improvement using the FedBal method that is considered as the future work of this research. We plan to develop a custom loss function for the semantic segmentation domain, and enhance our method as a service plugin that can be used on top of any federated learning algorithm to improve the robustness of the ML model.   

\section*{Acknowledgment}
We thank reviewers of the manuscript. The research was supported
by the Louisiana Board of Regents under grant number LEQSF(2016-19)-RDA-25.

\bibliography{mybibfile}

\begin{thebibliography}{10}
\expandafter\ifx\csname url\endcsname\relax
  \def\url#1{\texttt{#1}}\fi
\expandafter\ifx\csname urlprefix\endcsname\relax\def\urlprefix{URL }\fi
\expandafter\ifx\csname href\endcsname\relax
  \def\href#1#2{#2} \def\path#1{#1}\fi

\bibitem{ccinar2020machine}
Z.~M. {\c{C}}{\i}nar, A.~Abdussalam~Nuhu, Q.~Zeeshan, O.~Korhan, M.~Asmael,
  B.~Safaei, Machine learning in predictive maintenance towards sustainable
  smart manufacturing in industry 4.0, The Journal of Sustainability 12~(19)
  (2020) 8211.

\bibitem{hussain2022iot}
R.~F. Hussain, A.~Mokhtari, A.~Ghalambor, M.~A. Salehi, IoT for Smart
  Operations in the Oil and Gas Industry: From Upstream to Downstream, 1st
  Edition, 2022.

\bibitem{HUSSAIN2024479}
R.~F. Hussain, M.~A. Salehi, Resource allocation of industry 4.0 micro-service
  applications across serverless fog federation, Future Generation Computer
  Systems (FGCS) 154 (2024) 479--490.

\bibitem{hpcc20razin}
R.~F. Hussain, A.~Pakravan, M.~A. Salehi, Analyzing the performance of smart
  industry 4.0 applications on cloud computing systems, in: Proceedings of the
  22nd IEEE International Conference on High Performance Computing and
  Communications; IEEE 18th International Conference on Smart City; IEEE 6th
  International Conference on Data Science and Systems (HPCC/SmartCity/DSS),
  2020.

\bibitem{greenrazin19}
R.~F. Hussain, M.~Amini~Salehi, O.~Semiari, Serverless edge computing for green
  oil and gas industry, in: Proceedings of the IEEE Green Technologies
  Conference (GreenTech), 2019, pp. 1--4.

\bibitem{ums23}
T.~Chanikaphon, M.~Amini~Salehi, {UMS: Live Migration of Containerized Services
  across Autonomous Computing Systems}, in: Proceedings of the IEEE Global
  Communications Conference, GLOBECOM '23, 2023, pp. 467--472.

\bibitem{hussain2019federated}
R.~Hussain, M.~Amini, A.~Kovalenko, Y.~Feng, O.~Semiari, Federated edge
  computing for disaster management in remote smart oil fields, in: 2019 IEEE
  21st International Conference on High Performance Computing and
  Communications; IEEE 17th International Conference on Smart City; IEEE 5th
  International Conference on Data Science and Systems (HPCC/SmartCity/DSS),
  2019, pp. 929--936.

\bibitem{cadavid2020machine}
J.~P.~U. Cadavid, S.~Lamouri, B.~Grabot, R.~Pellerin, A.~Fortin, Machine
  learning applied in production planning and control: a state-of-the-art in
  the era of industry 4.0, Journal of Intelligent Manufacturing (2020) 1--28.

\bibitem{hossen2020object}
M.~I. Hossen, Y.~Tu, M.~F. Rabby, M.~N. Islam, H.~Cao, X.~Hei, An object
  detection based solver for google’s image recaptcha v2, in: 23rd
  International Symposium on Research in Attacks, Intrusions and Defenses
  ($\{$RAID$\}$ 2020), 2020, pp. 269--284.

\bibitem{singhai2021investigation}
R.~Singhai, R.~Sushil, An investigation of various security and privacy issues
  in internet of things, Materials Today: Proceedings (2021).

\bibitem{zob22}
S.~Zobaed, M.~Amini~Salehi, Privacy-preserving clustering of unstructured big
  data for cloud-based enterprise search solutions, Concurrency and
  Computation: Practice and Experience 34~(22) (2022) e7160.

\bibitem{zhou2020privacy}
C.~Zhou, A.~Fu, S.~Yu, W.~Yang, H.~Wang, Y.~Zhang, Privacy-preserving federated
  learning in fog computing, The Journal of Internet of Things 7~(11) (2020)
  10782--10793.

\bibitem{fedZehui}
J.~Kang, Z.~Xiong, D.~Niyato, S.~Xie, J.~Zhang, Incentive mechanism for
  reliable federated learning: A joint optimization approach to combining
  reputation and contract theory, IEEE Internet of Things Journal (2019).

\bibitem{duan2020self}
M.~Duan, D.~Liu, X.~Chen, R.~Liu, Y.~Tan, L.~Liang, Self-balancing federated
  learning with global imbalanced data in mobile systems, IEEE Transactions on
  Parallel and Distributed Systems 32~(1) (2020) 59--71.

\bibitem{mcmahan2017communication}
B.~McMahan, E.~Moore, D.~Ramage, S.~Hampson, B.~A. y~Arcas,
  Communication-efficient learning of deep networks from decentralized data,
  in: Artificial Intelligence and Statistics, 2017, pp. 1273--1282.

\bibitem{duan2019astraea}
M.~Duan, D.~Liu, X.~Chen, Y.~Tan, J.~Ren, L.~Qiao, L.~Liang, Astraea:
  Self-balancing federated learning for improving classification accuracy of
  mobile deep learning applications, in: 2019 IEEE 37th International
  Conference on Computer Design (ICCD), 2019, pp. 246--254.

\bibitem{mhaisen2020analysis}
N.~Mhaisen, A.~Awad, A.~Mohamed, A.~Erbad, M.~Guizani, Analysis and optimal
  edge assignment for hierarchical federated learning on non-iid data, arXiv
  preprint arXiv:2012.05622 (2020).

\bibitem{ye2020federated}
D.~Ye, R.~Yu, M.~Pan, Z.~Han, Federated learning in vehicular edge computing: A
  selective model aggregation approach, IEEE Access 8 (2020) 23920--23935.

\bibitem{veillon2019f}
V.~Veillon, C.~Denninnart, M.~A. Salehi, F-fdn: Federation of fog computing
  systems for low latency video streaming, in: 2019 IEEE 3rd International
  Conference on Fog and Edge Computing (ICFEC), 2019, pp. 1--9.

\bibitem{shokri2015privacy}
R.~Shokri, V.~Shmatikov, Privacy-preserving deep learning, in: Proceedings of
  the 22nd ACM SIGSAC conference on computer and communications security, 2015,
  pp. 1310--1321.

\bibitem{ma2020safeguarding}
C.~Ma, J.~Li, M.~Ding, H.~H. Yang, F.~Shu, T.~Q. Quek, H.~V. Poor, On
  safeguarding privacy and security in the framework of federated learning,
  IEEE network 34~(4) (2020) 242--248.

\bibitem{wang2019beyond}
Z.~Wang, M.~Song, Z.~Zhang, Y.~Song, Q.~Wang, H.~Qi, Beyond inferring class
  representatives: User-level privacy leakage from federated learning, in: IEEE
  INFOCOM 2019-IEEE Conference on Computer Communications, 2019, pp.
  2512--2520.

\bibitem{truex2019hybrid}
S.~Truex, N.~Baracaldo, A.~Anwar, T.~Steinke, H.~Ludwig, R.~Zhang, Y.~Zhou, A
  hybrid approach to privacy-preserving federated learning, in: Proceedings of
  the 12th ACM Workshop on Artificial Intelligence and Security, 2019, pp.
  1--11.

\bibitem{paillier1999public}
P.~Paillier, Public-key cryptosystems based on composite degree residuosity
  classes, in: International conference on the theory and applications of
  cryptographic techniques, 1999, pp. 223--238.

\bibitem{liu2020fedvision}
Y.~Liu, A.~Huang, Y.~Luo, H.~Huang, Y.~Liu, Y.~Chen, L.~Feng, T.~Chen, H.~Yu,
  Q.~Yang, Fedvision: An online visual object detection platform powered by
  federated learning, in: Proceedings of the AAAI Conference on Artificial
  Intelligence, Vol.~34, 2020, pp. 13172--13179.

\bibitem{pillutla2019robust}
K.~Pillutla, S.~M. Kakade, Z.~Harchaoui, Robust aggregation for federated
  learning, arXiv preprint arXiv:1912.13445 (2019).

\bibitem{liu2020experiments}
B.~Liu, B.~Yan, Y.~Zhou, Y.~Yang, Y.~Zhang, Experiments of federated learning
  for covid-19 chest x-ray images, arXiv preprint arXiv:2007.05592 (2020).

\bibitem{sheller2018multi}
M.~J. Sheller, G.~A. Reina, B.~Edwards, J.~Martin, S.~Bakas,
  Multi-institutional deep learning modeling without sharing patient data: A
  feasibility study on brain tumor segmentation, in: International MICCAI
  Brainlesion Workshop, 2018, pp. 92--104.

\bibitem{krestenitis2019oil}
M.~Krestenitis, G.~Orfanidis, K.~Ioannidis, K.~Avgerinakis, S.~Vrochidis,
  I.~Kompatsiaris, Oil spill identification from satellite images using deep
  neural networks, Remote Sensing 11~(15) (2019) 1762.

\bibitem{krestenitis2019early}
M.~Krestenitis, G.~Orfanidis, K.~Ioannidis, K.~Avgerinakis, S.~Vrochidis,
  I.~Kompatsiaris, Early identification of oil spills in satellite images using
  deep cnns, in: International Conference on Multimedia Modeling, 2019, pp.
  424--435.

\bibitem{yekeen2020novel}
S.~T. Yekeen, A.-L. Balogun, K.~B.~W. Yusof, A novel deep learning instance
  segmentation model for automated marine oil spill detection, ISPRS Journal of
  Photogrammetry and Remote Sensing 167 (2020) 190--200.

\bibitem{abraham2019novel}
N.~Abraham, N.~M. Khan, A novel focal tversky loss function with improved
  attention u-net for lesion segmentation, in: 2019 IEEE 16th International
  Symposium on Biomedical Imaging (ISBI 2019), 2019, pp. 683--687.

\bibitem{salehi2017tversky}
S.~S.~M. Salehi, D.~Erdogmus, A.~Gholipour, Tversky loss function for image
  segmentation using 3d fully convolutional deep networks, in: International
  workshop on machine learning in medical imaging, 2017, pp. 379--387.

\bibitem{ronneberger2015u}
O.~Ronneberger, P.~Fischer, T.~Brox, U-net: Convolutional networks for
  biomedical image segmentation, in: International Conference on Medical image
  computing and computer-assisted intervention, 2015, pp. 234--241.

\bibitem{ziller2021pysyft}
A.~Ziller, A.~Trask, A.~Lopardo, B.~Szymkow, B.~Wagner, E.~Bluemke, J.-M.
  Nounahon, J.~Passerat-Palmbach, K.~Prakash, N.~Rose, et~al., Pysyft: A
  library for easy federated learning, in: Federated Learning Systems, 2021,
  pp. 111--139.

\bibitem{bisong2019google}
E.~Bisong, Google colaboratory, in: Building Machine Learning and Deep Learning
  Models on Google Cloud Platform, 2019, pp. 59--64.

\end{thebibliography}

\end{document}